\documentclass{tlp}
\usepackage{amssymb}
\usepackage{amsmath}
\usepackage{stmaryrd}
\usepackage{pst-all}
\usepackage{xypic}
\usepackage{fancyvrb}
\usepackage{subfigure}
\usepackage{maybemath}
\usepackage{wasysym}
\usepackage{wmchrism}

\usepackage{bold-extra}
\usepackage{listings}
\usepackage{ifpdf}
\ifpdf
	\usepackage[pdftex]{graphicx}
	\usepackage{pdfpages}
	\usepackage[pdftex]{hyperref}
	\hypersetup{
	    pdfauthor={Jon Sneyers and Peter Van Weert and Tom Schrijvers and Leslie De Koninck},
	    pdftitle={\title},
	    breaklinks,
	    pdfstartview=FitH,
	    colorlinks,
	    linkcolor=black,
	    filecolor=black,
	    citecolor=black,
	    urlcolor=black
	}
	\usepackage[all]{hypcap}
	\usepackage{pdfsync} 
\else	
	\usepackage[dvips]{graphicx}
	\usepackage{url}
	\newcommand\href[2]{{\tt #2}}
\fi

\newcommand{\prob}{\mathit{prob}}

\newcommand{\chrism}{{\sf CHRiSM}}
\newcommand{\Chrism}{{\sf \bfseries CHRiSM}}
\newcommand{\omegachrism}{\ensuremath{\maybebm{\omega_t^{\mathtt{??}}}}}
\newcommand{\Stratchrism}{\ensuremath{\Omega_t^{\mathtt{??}}}}

\newcommand{\N}{\ensuremath{\mathbb{N}}}

\newcommand{\comment}[1]{}

        {\end{tabbing}\end{trivlist}}
\newbox\subfigbox
  

\newcommand{\noopsort}[1]{}

\newcommand{\Prog}{\mathcal{P}}

\newcommand{\HL}{\ensuremath{\mathcal{H}}}

\newcommand{\CT}{\mathcal{D}}
\newcommand{\BT}{\ensuremath{\CT_{\HL}}}
\newcommand{\musthold}{\ensuremath{\BT \models}}
\newcommand{\true}{\ensuremath{\mbox{\it true}}}
\newcommand{\false}{\ensuremath{\mbox{\it fail}}}

\newcommand{\exist}{\ensuremath{\bar \exists}}
\newcommand{\omegat}{\ensuremath{\maybebm{\omega_t}}}
\newcommand{\omegar}{\ensuremath{\maybebm{\omega_r}}}
\newcommand{\omegap}{\ensuremath{\maybebm{\omega_p}}}

\newcommand{\CHRv}{CHR$^{\vee}$}

\newcommand{\chr}{\mathit{chr}}
\newcommand{\vars}{\mathit{vars}}
\newcommand{\id}{\mathit{id}}

\newcommand{\state}{\ensuremath{\sigma}}
\newcommand{\States}{\ensuremath{\Sigma^{\mbox{\sc chr}}}}

\newcommand{\st}[5]{\ensuremath{\langle #1, #2, #3, #4 \rangle_{#5}}}

\newcommand{\ILetter}[1]{\ensuremath{\mathbb{#1}}}
\newcommand{\IG}{\ILetter{G}}

\newcommand{\IS}{\ILetter{S}}
\newcommand{\IB}{\ILetter{B}}
\newcommand{\IT}{\ILetter{T}}

\newcommand{\num}[2]{#1\##2}

\newcommand\transbase[3]{%
\newbox\WMTopBox \setbox\WMTopBox=\hbox{\scriptsize$#1$} %
\newbox\WMBottomBox \setbox\WMBottomBox=\hbox{\tiny$#2$} %
\newbox\WMArrowBox \setbox\WMArrowBox=\hbox{$#3{\hspace*{\wd\WMBottomBox\hskip1em}}$} %
\copy\WMArrowBox %
\hbox{\hskip-0.5\wd\WMArrowBox \hskip-0.5\wd\WMTopBox\hbox{\raise0.6em\copy\WMTopBox}}
\hskip-0.5\wd\WMTopBox \hskip0.5\wd\WMArrowBox\hskip-4pt\hskip-\wd\WMBottomBox\hbox{\lower0.5\ht\WMArrowBox\copy\WMBottomBox} 
\hskip4pt\hskip0.1em 
}
\newcommand\transaddstar{\hskip-0.2em\hbox{\raise0.3em\hbox{\scriptsize$*$}\hskip0.3em}}
\newcommand\nottrans{\hskip1em\not\hskip-1em}

\newcommand\trans[2]{\transbase{#1}{#2}{\wmexecpath}}
\newcommand\transx[2]{\transbase{#1}{#2}{\xrightarrow}}
\newcommand\transs[2]{\transbase{#1}{#2}{\wmexecpath}\transaddstar}
\newcommand\transsx[2]{\transbase{#1}{#2}{\xrightarrow}\transaddstar}
\newcommand\transsp[2]{\transbase{#1}{#2}{\wmexecpaths}}

\newcommand{\strat}{\transx{}{\xi,\Prog}}
\newcommand{\strata}{\ensuremath{\transx{}{\xi_1,\Prog}}}
\newcommand{\stratb}{\ensuremath{\transx{}{\xi_2,\Prog}}}

\newcommand{\Strat}{\ensuremath{\Omega}}

\newtheorem{theorem}{Theorem}[section]

\newtheorem{definition}[theorem]{Definition}

\begin{document}
\submitted{5 February 2010}
\revised{21 April 2010}
\accepted{20 March 2010}

\title{
CHR(PRISM)-based Probabilistic Logic Learning} 

\author[Jon Sneyers et.al.]{JON SNEYERS, WANNES MEERT, JOOST VENNEKENS\\
Dept.\ of Computer Science, K.U.Leuven, Belgium\\
\email{\{jon.sneyers,wannes.meert,joost.vennekens\}@cs.kuleuven.be}
\and YOSHITAKA KAMEYA and TAISUKE SATO\\
Tokyo Institute of Technology, Japan \\
\email{\{kameya,sato\}@mi.cs.titech.ac.jp}
}
\comment{ 
\author{     Jon Sneyers \inst{1} 
        \and Wannes Meert \inst{1} 
        \and Joost Vennekens \inst{1}
        \and \\
        Yoshitaka Kameya \inst{2} 
        \and Taisuke Sato \inst{2}}
\institute{
        K.U.Leuven, Belgium \\
        \email{\{jon.sneyers,wannes.meert,joost.vennekens\}@cs.kuleuven.be}
        \and
        Tokyo Institute of Technology, Japan \\
        \email{\{kameya,sato\}@mi.cs.titech.ac.jp}
}	
}
\maketitle

\begin{abstract}
PRISM 
is an extension of
Prolog with probabilistic predicates and built-in support
for expectation-maximization learning.
Constraint Handling Rules (CHR) is a high-level programming language
based on multi-headed multiset rewrite rules.

In this paper, we introduce a new probabilistic logic formalism, called \chrism{},
based on a combination of CHR and PRISM.
It can be used for high-level rapid prototyping of 
complex statistical models by means of ``chance rules''.
The underlying PRISM system can then be used for several probabilistic
inference tasks, including probability computation and parameter learning.
We define the \chrism{} language in terms of syntax and operational semantics,
and illustrate it with examples.
We define the notion of ambiguous programs and define a distribution semantics
for unambiguous programs.
Next, we describe an implementation of \chrism{}, based on CHR(PRISM).
We discuss the relation between \chrism{} and other
probabilistic logic programming languages, in particular PCHR.
Finally, we identify potential application domains.
\end{abstract}

\section{Introduction}

Constraint Handling Rules \cite{fru_chr_2008,newsurvey}
is a high-level language extension based on multi-headed rules.
Originally, CHR was designed as a special-purpose language to implement
constraint solvers, but in recent years it has matured into a general purpose
programming language. 
%
Being a language \emph{extension}, CHR is implemented on top of an
existing programming language, which is called the \emph{host language}.
An implementation of CHR in host language $X$ is called CHR($X$).
For instance, several CHR(Prolog) systems are available.

PRISM (PRogramming In Statistical Modeling) is a probabilistic extension of
Prolog \cite{prism}. It supports
several probabilistic inference tasks, including sampling,
probability computation, and expectation-maximization (EM)
learning.

In this paper, we construct a new formalism, called \chrism{}
--- short for {\bf CH}ance {\bf R}ules {\bf i}nduce {\bf S}tatistical {\bf M}odels. 
It is based on CHR(PRISM) and it combines the advantages of CHR and those of PRISM.
Like CHR, \chrism{} is a very concise and expressive programming language.
Like PRISM, \chrism{} has built-in support for several probabilistic inference tasks.
Furthermore, since \chrism{} is implemented as a translation to CHR(PRISM) --- which itself is
translated to PRISM and ultimately Prolog --- \chrism{} rules can be freely mixed
with CHR rules and Prolog clauses.

This paper is based on an earlier workshop paper \cite{sney_meert_vennekens_chrism_chr09}.
Although it is mostly self-contained, some familiarity
with CHR and PRISM is recommended.

We use $\uplus$ for multiset union, $\subsetpluseq$ for multiset subset,
and $\exist_{A} B$ to denote 
$\exists x_1, \ldots, x_n: B$, with $\{x_1,\ldots,x_n\} = \vars(B) \setminus \vars(A)$,
where $\vars(A)$ are the (free) variables in $A$;
if $A$ is omitted it is empty
(so $\exist B$ denotes the existential closure of $B$).

\section{Syntax and Semantics of \Chrism{}}

In this section we define \chrism{}. 
The syntax is defined in Section~\ref{sec:syntax}
and the (abstract) operational semantics is defined in Section~\ref{sec:semantics}.
Finally, in Section~\ref{sec:observations} the notion of \emph{observations}
is introduced.


\subsection{Syntax and Informal Semantics}
\label{sec:syntax}
A \chrism{} program $\Prog$ consists of a sequence of \emph{chance rules}.
Chance rules rewrite a multiset $\IS$ of data elements, which are called (\chrism{}) \emph{constraints}
(mostly for historical reasons). Syntactically, a constraint {\tt c(X$_1$,..,X$_n$)}
looks like a Prolog predicate:
it has a functor {\tt c} of some arity $n$ and arguments {\tt X$_1$,..,X$_n$}
which are Prolog terms.
The multiset $\IS$ of constraints is called the \emph{constraint store} or just \emph{store}.
The initial store is called the \emph{query} or \emph{goal},
the final store (obtained by exhaustive rule application) is called the \emph{answer} or \emph{result}.

\paragraph{Chance rules.}
A chance rule is of the following form:
\begin{verbatim}
   P ?? Hk \ Hr <=> G | B.
\end{verbatim}
where {\tt P} is a probability expression (as defined below),
{\tt Hk} is a conjunction of (kept head) constraints,
{\tt Hr} is a conjunction of (removed head) constraints,
{\tt G} is a guard condition (a Prolog goal to be satisfied),
and {\tt B} is the body of the rule.
If {\tt Hk} is empty, the rule is called a \emph{simplification} rule and
the backslash is omitted; if {\tt Hr} is empty, the rule is
called a \emph{propagation} rule and it is written as ``{\tt P ?? Hk ==> G | B}''.
If both {\tt Hk} and {\tt Hr} are non-empty, the rule is called a \emph{simpagation} rule.
The guard {\tt G} is optional; if it is removed, the ``{\tt |}'' is also removed.
The body {\tt B} is recursively defined as a conjunction of \chrism{} constraints, Prolog goals,
and probabilistic disjunctions (as defined below) of bodies.

Intuitively, the meaning of a chance rule is as follows:
If the constraint store $\IS$ contains elements that match with the head of the rule
(i.e. if there is a matching substitution $\theta$ such that 
$(\theta(\mathtt{Hk}) \uplus \theta(\mathtt{Hr})) \subsetpluseq \IS$),
and furthermore, the guard {\tt G} is satisfied, then we can consider rule application.
The subset of $\IS$ that corresponds to the head of the rule is called a rule \emph{instance}.
Depending on the probability expression {\tt P}, the rule instance is either ignored
or it actually leads to a rule application. 
Every rule instance may only be considered once.

Rule application has the following effects:
the constraints matching {\tt Hr} are removed from the constraint store,
and then the body {\tt B} is executed, that is, Prolog goals are called
and \chrism{} constraints are added into the 
store.

\paragraph{Probability expressions.}
A probability expression {\tt P} is one of the following:
\begin{itemize}
\item A number from 0 to 1, indicating the probability that the rule fires. 
        A rule of the form {\tt 1 ?? ...} corresponds to a regular CHR rule; 
        the ``{\tt 1 ??}'' may be dropped. A rule of the form {\tt 0 ?? ...} is never applied.
\item An expression of the form {\tt eval(E)}, where {\tt E} is an arithmetic
        expression (in Prolog syntax).
        It should be ground when the rule is considered
        (otherwise a runtime instantiation error occurs).
        The evaluated expression indicates the probability that the rule fires.
\item An experiment name. This is a Prolog term which should be ground 
        when the rule is considered.
        The probability distribution is unknown.
        Initially, unknown probabilities are set to a uniform distribution 
        ({\tt 0.5} in the case of rule probabilities). 
        They can be changed manually using PRISM's {\tt set\_sw}/2 builtin,
        or automatically using PRISM's EM-learning algorithm.
        
        The arguments of the experiment name can include \emph{conditions},
        which are of the form ``{\tt cond C}''. Such arguments are evaluated at
        runtime and replaced by either ``{\tt yes}'' or ``{\tt no}'', depending
        on whether {\tt call(C)} succeeded or failed.

        These conditions 
        are just syntactic sugar,
        so we may ignore them w.l.o.g. 
        For example, the rule
        ``{\tt foo(cond~A>B) ?? c(A,B) <=> d}''
        is syntactic sugar for
        ``{\tt foo(X) ?? c(A,B) <=> (A>B -> X=yes ; X=no) | d}''.
\item Omitted (so the rule starts with ``{\tt ??}''): 
        this is a shorthand for a fresh zero-arity experiment name.
\end{itemize}

\paragraph{Probabilistic disjunction.}
The body {\tt B} of a \chrism{} rule 
may contain
probabilistic disjunctions. 
There are 
two styles: 
\begin{itemize}
\item LPAD-style probabilistic disjunctions \cite{lpad_iclp04}
        of the form 
        ``{\tt D1:P1 ; ... ; Dn:Pn}'',
        where  a disjunct {\tt D$i$} is chosen with probability {\tt P$i$}.
        The probabilities should sum to 1
        (otherwise a compile-time error occurs).
\item \chrism{}-style probabilistic disjunctions of the form
        ``{\tt P ?? D1 ; ... ; Dn}'',
         where {\tt P} is an experiment name determining
        the probability distribution.
\end{itemize}
The LPAD-style probabilistic disjunctions can be seen as a special case
of \chrism{}-style disjunctions for which the experiment name is implicit
and the distribution is given and fixed.
Unlike \CHRv{} disjunctions, which create a choice point, 
both kinds of probabilistic disjunctions are \emph{committed-choice}:
once a disjunct is chosen, the choice is not undone later. 
However, when later on in a derivation the same experiment is sampled again,
the result can of course be different.

\subsection{Operational Semantics}
\label{sec:semantics}

\begin{figure}
\fbox{
\begin{minipage}{0.97\textwidth}
\begin{description} 
\item[\textbf{1. Fail.}]
$
        \st{\{b\} \uplus \IG}{\IS}{\IB}{\IT}{n}
        \ \trans{1}{\Prog}\ 
        \false
$\\
where $b$ is a built-in (Prolog) constraint and
$\musthold \lnot \exist (\IB \land b)$.

\item[\textbf{2. Solve.}]
$
        \st{\{b\} \uplus \IG}{\IS}{\IB}{\IT}{n}
        \ \trans{1}{\Prog}\ 
        \st{\IG}{\IS}{b \land \IB}{\IT}{n}
$\\
where $b$ is a built-in (Prolog) constraint and
$\musthold \exist (\IB \land b)$.


\item[\textbf{3. Introduce.}]
$
        \st{\{c\} \uplus \IG}{\IS}{\IB}{\IT}{n}
        \ \trans{1}{\Prog}\ 
        \st{\IG}{\{\num{c}{n}\} \cup \IS}{\IB}{\IT}{n+1}
$\\
where $c$ is a \chrism{} constraint.


\item[\textbf{4. Probabilistic-Choice.}]
$
        \st{\{d\} \uplus \IG}{\IS}{\IB}{\IT}{n}
        \ \trans{p_i}{\Prog}\ 
        \st{\{d_i\} \uplus \IG}{\IS}{\IB}{\IT}{n}
$\\
where $d$ is a probabilistic disjunction of the form
{\tt $d_1$:$p_1$ ; $\ldots$ ; $d_k$:$p_k$} or of the form \linebreak[4]
{\tt P ?? $d_1$ ; $\ldots$ ; $d_k$}, where the probability
distribution given by {\tt P} assigns the probability $p_i$ to
the disjunct $d_i$.

\item[\textbf{5. Maybe-Apply.}]
\hfill
$
        \st{\IG}{H_1 \uplus H_2 \uplus \IS}{\IB}{\IT}{n}
 5;3~5;3~       \ \trans{1-p}{\Prog}\ 
        \st{\IG}{H_1 \uplus H_2 \uplus \IS}{\IB}{\IT \cup \{h\}}{n}
$\\
{\textcolor{white}{foo}}\hfill  
$
        \st{\IG}{H_1 \uplus H_2 \uplus \IS}{\IB}{\IT}{n}
        \ \trans{p}{\Prog}\ 
        \st{B \uplus \IG}{H_1 \uplus \IS}{\theta \land \IB}{\IT \cup \{h\}}{n}
$\\
where the $r$-th rule of $\Prog$ is of the form 
{\tt P ?? $H'_1\ \backslash\ H'_2$ <=> G | B},\\
$\theta$ is a matching substitution such that
$\chr(H_1) = \theta(H'_1)$ and $\chr(H_2) = \theta(H'_2)$,
$ h = (r,\id(H_1),\id(H_2)) \not\in \IT $,
and $\musthold \IB \rightarrow \exist_\IB (\theta \wedge G)$.
If {\tt P} is a number, then $p= \mathtt{P}$. Otherwise $p$ is the probability
assigned to the success branch of {\tt P}.
\end{description}
\end{minipage}
}
\caption{
Transition relation $\trans{}{\Prog}$
of the abstract operational semantics $\omegachrism$ of \chrism{}. 
}
\label{fig:omegachrism}
\end{figure}

The abstract operational semantics of a \chrism{} program $\Prog$ is given by
a state-transition system that resembles the abstract operational semantics
\omegat{} of CHR \cite{newsurvey}.
The execution states are defined analogously,
except that we additionally define a unique 
failed execution state which is denoted by ``$\false$''
(because we don't want to distinguish between different failed states).
We use the symbol \omegachrism{} to refer to the abstract operational semantics of \chrism{}.

\begin{definition}[identified constraint]
An \emph{identified} constraint $\num{c}{i}$ is a \chrism{} constraint 
$c$ associated with some unique integer $i$. 
This number serves to differentiate between copies of the same constraint.
We introduce the functions $\chr(\num{c}{i}) = c$ and $\id(\num{c}{i}) = i$,
and extend them to sequences and sets 
in the obvious manner, e.g., $\id(S) = \{ i |\num{c}{i} \in S\}$.
\end{definition}

\begin{definition}[execution state]
An \emph{execution state} $\sigma$ is a tuple
$ \langle \IG, \IS, \IB, \IT \rangle_n$.
The \emph{goal} $\IG$ is a multiset of constraints to be rewritten to solved form.
The \emph{store} $\IS$ is a set of \emph{identified} 
constraints that can be matched with rules in the program $\Prog$.
Note that $\chr(\IS)$ is a multiset although $\IS$ is a set.
The \emph{built-in store} $\IB$ is the conjunction of all 
Prolog goals that have been called so far.
%
The \emph{history} $\IT$ is a set of tuples, 
each recording the identifiers of the \chrism{} constraints that fired a rule
and the rule number.
The history is used to prevent trivial non-termination:
a rule instance is allowed to be considered only once.
%
Finally, the counter $n \in \N$ represents the next free identifier.
\end{definition}

We use $\sigma, \sigma_0, \sigma_1, \ldots$ to denote execution states and
$\States$ to denote the set of all execution states.
We use $\BT$ to denote the theory defining the host language (Prolog) 
built-ins and predicates used in the \chrism{} program. 
For a given CHR program $\Prog$, the transitions are defined by
the binary relation 
$\trans{}{\Prog} \subset \States \times \States$
shown in Figure~\ref{fig:omegachrism}.
Every transition is annotated with a probability.

Execution proceeds by exhaustively applying the transition rules, starting
from an initial state (root) of the form 
$\langle Q, \emptyset, \true, \emptyset \rangle_0$
and performing a random walk in the directed acyclic graph
defined by the transition relation
$\trans{}{\Prog}$, until a leaf node is reached, which is called a final state.
We consider only terminating programs (finite transition graphs).
Given a path from an initial state to the state $\sigma$, we define the probability
of $\sigma$ to be the product of the probabilities along the path.
We use $\sigma_0 \transs{p}{\Prog} \sigma_k$
to denote a series of $k \geq 0$ transitions 
$$\sigma_0\ \trans{p_1}{\Prog}\ \sigma_1\ \trans{p_2}{\Prog}\ \sigma_2\ \trans{p_3}{\Prog}\ \ldots\ \trans{p_k}{\Prog}\ \sigma_k$$
where $p = \prod_{i=1}^k p_i$ if $k>0$ and $p=1$ otherwise.
If $\sigma_0$ is an initial state and $\sigma_k$ is a final state,
then we call such a series of transitions a \emph{derivation}
of probability $p$. We define a function $\prob$ to give the probability of a derivation:
$\prob(\sigma_0 \transs{p}{\Prog} \sigma_k) = p$.

Note that if all rule probabilities are 1 and the program contains
no probabilistic disjunctions --- i.e. if the \chrism{} program is actually
just a regular CHR program --- then the \omegachrism{} semantics
boils down
to the \omegat{} semantics of CHR.




\subsection{Full and Partial Observations}
\label{sec:observations}

A full observation {\tt Q <==> A} denotes that there exist a 
series of probabilistic choices such that
a derivation starting with query {\tt Q} results in the answer {\tt A}.
A partial observation {\tt Q ===> A} denotes that an answer for query {\tt Q}
contains at least {\tt A}: in other words, {\tt Q ===> A} holds iff {\tt Q <==> B} with {\tt A $\subsetpluseq$ B}.

\begin{definition}[observation]
A \emph{full observation} is of the form {\tt Q <==> A},
where {\tt Q} and {\tt A} are conjunctions of 
constraints.
Given a 
program $\Prog$, a full observation refers to 
derivations of the form 
$$\langle Q, \emptyset, \true, \emptyset \rangle_0\ \transs{p}{\Prog}\ 
\langle \emptyset, A', \IB, \IT \rangle_n \nottrans\trans{}{\Prog} $$
such that $\mathtt{A} = \chr(A')$.
A \emph{partial observation} is of the form  {\tt Q ===> A}.
It refers to derivations of the above form,
such that $\mathtt{A} \subsetpluseq \chr(A')$.
\end{definition}

We also allow ``negated'' \chrism{} constraints in the right hand side:\\
{\tt Q~===>~A,$\sim$N} is a shorthand for {\tt Q <==> B} with {\tt A $\subsetpluseq$ B}
and {\tt N $\not\subsetpluseq$ B $\setminus$ A}.

\medskip

\noindent The following PRISM built-ins can be used to query a \chrism{} program:
\begin{itemize}
\item {\tt sample Q} : probabilistically execute the query {\tt Q};
\item {\tt prob Q <==> A} : compute the probability that {\tt Q <==> A} holds, i.e. the chance that
        the choices are such that query {\tt Q} results in answer {\tt A};
\item {\tt prob Q ===> A} : compute the probability that an answer for {\tt Q} contains {\tt A};
\item {\tt learn(L)} : perform EM-learning based on a list {\tt L} of observations
\end{itemize}

\noindent In observation lists, the syntax ``{\tt $n$ times $X$}'' or ``{\tt count($X$,$n$)}'' 
can be used to denote that observation $X$ occurred $n$ times. 
This is simply a shorthand for repeating the same observation 
($X$) a number of times ($n$).

\section{Example programs}

As a first toy example, consider the following \chrism{} program for tossing a coin:
\begin{verbatim}
   toss <=> head:0.5 ; tail:0.5.
\end{verbatim}
The query ``{\tt sample toss}'' results in ``{\tt head}'' or ``{\tt tail}'',
with 50\% chance each. 
The query ``{\tt sample toss,toss}'' has four possible outcomes, each
with 25\% chance:
``{\tt head,head}'', ``{\tt head,tail}'', 
``{\tt tail,head}'', and ``{\tt tail,tail}''.

Note that observations are not sensitive to the order in which the result is given.
As a result, the query ``{\tt prob toss,toss <==> head,tail}'' 
returns a 
probability of 50\%, because the outcome ``{\tt tail,head}'' also
matches the observation.

\subsection{Rock-paper-scissors}
\label{sec:rsp}
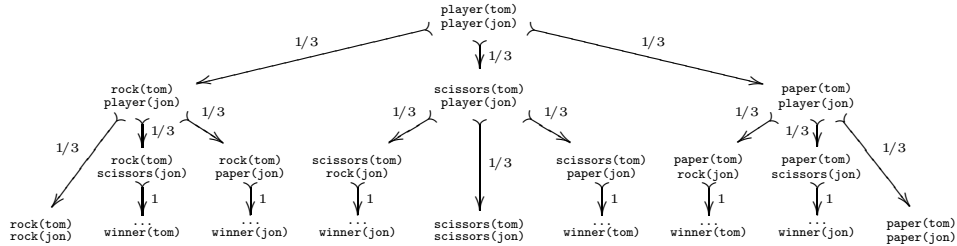
\begin{figure}
\begin{center}
\tiny
\ttfamily
\xyoption{tips}
\newcommand{\myar}{\ar@{*{\UseTips\dir{>}}-*{\NoTips\dir{>}}}}
\newcommand{\onethird}{1\!/3}
\xymatrix@R=1.5em@C=0.5em{
\SelectTips{cm}{11}
     & &   & & *++\txt{player(tom)\\player(jon)} 
     \myar[dlll]_{\onethird}     \myar[d]^{\onethird} \myar[drrr]^{\onethird}                         & & & & \\
 & *++\txt{rock(tom)\\player(jon)}  
        \myar@<-1ex>[ddl]_{\onethird} \myar[d]^{\onethird} \myar@<1ex>[dr]^{\onethird}
                      &  & & *++\txt{scissors(tom)\\player(jon)}  
                                \myar@<-1ex>[dl]_{\onethird} \myar[dd]^{\onethird} \myar@<1ex>[dr]^{\onethird}
                                     &   & & *++\txt{paper(tom)\\player(jon)}  
                                                \myar@<-1ex>[dl]_{\onethird} \myar[d]_{\onethird} \myar@<1ex>[ddr]^{\onethird}    &   \\
 & *+\txt{rock(tom)\\scissors(jon)} \myar[d]^1 & *+\txt{rock(tom)\\paper(jon)} \myar[d]^1& 
*+\txt{scissors(tom)\\rock(jon)} \myar[d]^1 &  & *+\txt{scissors(tom)\\paper(jon)} \myar[d]^1 & 
*+\txt{paper(tom)\\rock(jon)} \myar[d]^1 & *+\txt{paper(tom)\\scissors(jon)} \myar[d]^1 & \\
*+\txt{rock(tom)\\rock(jon)}& *+\txt{...\\winner(tom)} & *+\txt{...\\winner(jon)} & *+\txt{...\\winner(jon)} & *+\txt{scissors(tom)\\scissors(jon)} & *+\txt{...\\winner(tom)} & *+\txt{...\\winner(tom)} & *+\txt{...\\winner(jon)} & *+\txt{paper(tom)\\paper(jon)}
}
\end{center}
\caption{A derivation tree for the rock-paper-scissors example.}
\label{fig:example}
\end{figure}

Consider the following \chrism{} program
simulating 
``rock-paper-scissors'' players:
\begin{verbatim}
   player(P) <=> choice(P) ?? rock(P) ; scissors(P) ; paper(P).
   rock(P1), scissors(P2) ==> winner(P1).
   scissors(P1), paper(P2) ==> winner(P1).
   paper(P1), rock(P2) ==> winner(P1).
\end{verbatim}
We assume that each player has his own fixed probability distribution 
for choosing between rock, scissors, and paper. 
This is denoted by using {\tt choice(P)} as the 
probability expression for the choice in the first
rule: the probability distribution depends on the value of {\tt P} and thus every player
has his own distribution.
However, these distributions are not known to us.  
By default, the unknown probability distributions for, say, {\tt tom} and
{\tt jon} are therefore both set to the uniform distribution, 
which implies, among other things, that each player should win one third of the time
(cf.\ Figure~\ref{fig:example}).
\newcommand{\userinput}[1]{#1}
Here is a possible interaction: 
\begin{Verbatim}[commandchars=\\\{\}]
?- \userinput{sample player(tom),player(jon)}
player(tom),player(jon) <==> rock(jon),rock(tom).
?- \userinput{sample player(tom),player(jon)}
player(tom),player(jon) <==> rock(jon),paper(tom),winner(tom).
?- \userinput{prob player(tom),player(jon) ===> winner(tom)}
Probability of player(tom),player(jon)===>winner(tom) is: 0.333333
\end{Verbatim}

Now suppose that we watch 100 games, and want to use our observations to obtain a better model of the playing style of both players.  If we can fully observe these games, then this is easy: we can just use the frequency with which each player played rock, paper or scissors as an estimate for the probability of him making that particular move.  The situation becomes more difficult, however, if the games are only partly observable.  For instance, suppose that we do not know which moves the players made, but are only told the final scores: {\tt tom} won 50 games, {\tt jon} won 20, and 30 games
were a tie.  Deriving estimates for the probabilities of  individual moves from this information is less straightforward.  For this reason, PRISM comes with a built-in implementation of the EM-algorithm for performing parameter estimation in the presence of missing information \cite{Kameya00a:proc}.  We can use this algorithm to
find plausible corresponding distributions:

\begin{Verbatim}[commandchars=\\\{\},codes={\catcode`$=3}]
| ?- \userinput{learn([ (50 times  player(tom),player(jon) ===> winner(tom)),}
     \userinput{        (20 times  player(tom),player(jon) ===> winner(jon)),}
\userinput{(30 times  player(tom),player(jon) ===> $\sim$winner(tom),$\sim$winner(jon))])}
\end{Verbatim}
The PRISM built-in {\tt show\_sw} shows the 
learned
probability distributions,
which do indeed (approximately) lead to the observation frequencies, e.g.:
\begin{Verbatim}[commandchars=\\\{\}]
| ?- \userinput{show_sw}
Switch choice(jon): 1 (p: 0.60057) 2 (p: 0.06536) 3 (p: 0.33406)
Switch choice(tom): 1 (p: 0.08420) 2 (p: 0.20973) 3 (p: 0.70605)
| ?- \userinput{prob player(tom),player(jon) ===> winner(tom)}
Probability of player(tom),player(jon)===>winner(tom) is: 0.499604
\end{Verbatim}

\subsection{Random graphs}
\label{sec:rgraph}
Suppose we want to generate a random directed graph, given its nodes.
The following rule generates every possible directed edge with
probability 50\%:
\begin{verbatim}
   0.5 ?? node(A), node(B) ==> edge(A,B).
\end{verbatim}
The above rule generates dense graphs; if we want to get
a sparse graph, say with an average (out-)degree of 3, we can use
the following rule. The auxiliary constraint {\tt nb\_nodes$(n)$} contains
the total number of nodes $n$; the probability of the rule is such that
each of the $n(n-1)$ possible edges is generated with probability
$3/(n-1)$, so on average it generates $3n$ edges:
\begin{verbatim}
   eval(3/(N-1)) ?? nb_nodes(N), node(A), node(B) ==> edge(A,B).
\end{verbatim}

\subsection{Bayesian networks}\label{sec:bn}

Bayesian networks are one of the most widely used kinds of probabilistic models. 
A classical example \cite{pearl:firstbook} of a Bayesian network is that describing 
the following alarm system.
Suppose there is some probability that there is a burglary, and also that
there is some probability that an earthquake happens.
The probability that the alarm goes off depends on whether those events happen.
Also, the probability that John calls the police depends on whether the alarm
went off, and similarly for the probability that Mary calls.

\noindent
This Bayesian network can be described in \chrism{} in a straightforward way:
\begin{verbatim}
   go ==> ?? burglary(yes) ; burglary(no).
   go ==> ?? earthquake(yes) ; earthquake(no).
   burglary(B), earthquake(E) ==> B,E ?? alarm(yes) ; alarm(no).
   A ?? alarm(A) ==> johncalls.
   A ?? alarm(A) ==> marycalls.
\end{verbatim}
The probability distributions can be estimated given full observations (e.g., {\tt go <==> go, burglary(no), earthquake(yes), alarm(yes), marycalls.}), or given partial observations (e.g., {\tt\catcode`$=3\catcode`\_=8 go ===> johncalls, $\sim$marycalls.}).

In this way, each Bayesian network can be represented in \chrism{}.
We can derive the same information from it as can be derived from the network itself.

\comment{
\subsection{Chemical reactions}
\label{sec:chem}
Suppose we want to model the chemical reaction of molecules.

0.1 ?? h2, h2, o2 <=> h2o, h2o.
}

%
%

\section{Ambiguity and a Distribution Semantics for \Chrism{}}

In addition to the very nondeterministic abstract operational semantics \omegachrism{},
we can also define more deterministic instantiations of \omegachrism{},
just like \omegar{} and \omegap{} are instantiations of \omegat{}
(see also \cite{sney_fru_general_chr_chr08}).
In the current implementation of \chrism{} we use the ``refined semantics of \chrism{}'',
defined analogously to \cite{duck_stuck_garc_holz_refined_op_sem_iclp04}.
Of course \chrism{} can also be given a ``priority semantics''
\cite{dekoninck_schr_demoen_chrrp_ppdp07}
in order to get a more intuitive mechanism for execution control.

\subsection{Instantiations of \omegachrism{}}
\label{sec:instantiations}

Any \chrism{} system uses a (computable) execution strategy in the sense
of \cite{sney_fru_general_chr_chr08}. 
Note that in \cite{sney_fru_general_chr_chr08},
an execution strategy completely fixes the derivation for a given input goal.
In the context of \chrism{} this is no longer the case because of the probabilistic choices.
However, we may assume that the derivation is fixed if the same choices are made.
In other words, the only choice is in the probabilistic choices inside the transitions
``{\bf Probabilistic Choice}'' and ``{\bf Maybe-Apply}''; there is no nondeterminism
in choosing which \omegachrism{} transition to apply next.

\begin{definition}[execution strategy]
An \emph{execution strategy} fixes the non-probabilistic choices during an \omegachrism{} derivation.
Formally, $\strat$  is an execution strategy for a program $\Prog$ if 
$\strat \subseteq \trans{}{\Prog}$ and
for every execution state $\state \in \States$, the set $S$ of all transitions of the form $\state \strat \state'$
corresponds to at most one of the five types of transitions of \omegachrism{},
that is, either 
\begin{itemize}
\item $S = \emptyset$ and no \omegachrism{} transition is applicable;
\item or $S$ is a singleton corresponding to a {\bf Fail}, {\bf Solve} or
        {\bf Introduce} transition;
\item or $S$ is a set of transitions corresponding to the {\bf Probabilistic-Choice} transition
        for one specific disjunction;
\item or $S$ is a set of transitions corresponding to the {\bf Maybe-Apply} transition
        for one specific rule instantiation.
\end{itemize}        
\end{definition}
It follows from this definition that for non-final states $\state$,
the sum of the probabilities of all transitions from $\state$ is one
under any execution strategy.
We use $\sigma_0 \transsx{p}{\xi,\Prog} \sigma_k$
to denote a series of $k \geq 0$ transitions 
$$\sigma_0\ \transx{p_1}{\xi,\Prog}\ \sigma_1\ \transx{p_2}{\xi,\Prog}\ \sigma_2\ \transx{p_3}{\xi,\Prog}\ \ldots\ \transx{p_k}{\xi,\Prog}\ \sigma_k$$
where $p = \prod_{i=1}^k p_i$ if $k>0$ and $p=1$ otherwise, as before.

\begin{definition}[strategy class]
A \emph{strategy class} $\Strat(\Prog)$ is a set of 
execution strategies for $\Prog$.
The strategy class $\Stratchrism(\Prog)$ is the set of \emph{all} execution strategies for $\Prog$.
\end{definition}

\subsection{Distribution Semantics}
\label{sec:distribution_semantics}

Firstly, we define equivalence of execution states.
We use a definition based on \cite{equivalent_states} but adapted to our needs.
Intuitively, we say two states are equivalent if the constraint stores are equal
and the built-in stores are equivalent; we do not care about identifiers and
propagation histories.
\begin{definition}[equivalent states]
Equivalence between execution states is the smallest equivalence relation $\equiv$ 
s.t.:
\begin{enumerate}
\item $\st{\IG}{\IS}{{x = t \land \IB}}{\IT}{n} \equiv \st{\IG}{{\IS[x/t]}}{{x = t \land \IB}}{\IT'}{n'}$ 
\item $\st{\IG}{\IS}{{x = t \land \IB}}{\IT}{n} \equiv \st{{\IG[x/t]}}{\IS}{{x = t \land \IB}}{\IT'}{n'}$ 
\item $\st{\IG}{\IS}{\IB}{\IT}{n} \equiv \st{\IG}{\IS'}{\IB}{\IT'}{n'}$ 
        if $\chr(\IS) = \chr(\IS')$
\item $\st{\IG}{\IS}{\IB}{\IT}{n} \equiv \st{\IG}{\IS}{\IB'}{\IT}{n}$ 
        if $\musthold \exist_{\IG,\IS} \IB \leftrightarrow \exist_{\IG,\IS} \IB'$
\end{enumerate}
\end{definition}

We now define the probability of getting some result (given an execution strategy)
as the sum of the probabilities of ending up in a final state equivalent with it:
\begin{definition}[observation probability]
Given a program $\Prog$ and an execution strategy $\strat \in \Stratchrism(\Prog)$,
we write
 $$\sigma_i \transsp{p_\mathit{tot}}{\xi,\Prog} \sigma_f$$
if
$\sigma_f$ is a final state and
$p_\mathit{tot} = \sum_{d \in D} \prob(d)$
where $D = \{ \sigma_i \transsx{p}{\xi,\Prog} \sigma'_f \ |\ \sigma'_f \equiv \sigma_f \}$.
We say that $p_\mathit{tot}$ is the probability of
observing the result $\sigma_f$ for the query $\sigma_i$.
\end{definition}

\subsection{Ambiguity}
\label{sec:ambiguity}
Some programs are \emph{ambiguous} in the sense that they do not define
a unique probability distribution over the possible end states.
Consider the following example:
\begin{verbatim}
   0.5 ?? a <=> b.
   0.5 ?? a <=> c.
\end{verbatim}
Suppose the query is ``{\tt a}''. If we use an execution strategy that
starts with the first rule, then with 50\% chance this rule is applied
and we get the final result ``{\tt b}'', with 50\% chance the second rule is considered 
resulting in ``{\tt c}'' with a probability of 25\%, and when no rule is applied
 the result is ``{\tt a}'' with a probability of 25\%.
However, if we use an execution strategy that considers the second rule first,
then we get a different distribution: ``{\tt c}'' has a probability of 50\%, 
and ``{\tt b}'' a probability of 25\%.


A program is unambiguous if the probability of an observation does not depend on the execution strategy.
The program in the above example is ambiguous in general, but it is unambiguous w.r.t. the refined
strategy class. Under the refined semantics, the first rule is always considered first, 
thus the above program defines only the first 
probability distribution on final states.
In general, we define ambiguity w.r.t. a strategy class --- if the strategy class is omitted,
we assume it is the most general strategy class corresponding to all execution strategies
that instantiate \omegachrism{}.

\begin{definition}[unambiguous program]
A \chrism{} program $\Prog$ is unambiguous (w.r.t. a strategy class $\Strat$) if,
for all states $\state_i, \state_f \in \States$ and all execution strategies
$\strata, \stratb \in \Strat$, 
we have:

if $\state_i \transsp{p_1}{\xi_1,\Prog} \state_f$ 
and $\state_i \transsp{p_2}{\xi_2,\Prog} \state_f$ 
then $p_1 = p_2$.
\end{definition}

The distribution semantics (w.r.t. strategy class $\Strat$)
of an unambiguous (w.r.t. $\Strat$) \chrism{} program 
is defined for every query $Q$
as the probability distribution over the equivalence classes of
final states of derivations (of $\Strat$).

Without specification of an execution strategy, ambiguous \chrism{} programs 
do not have a well-defined meaning --- they
don't define a unique probability distribution over the final states,
but \emph{several} distributions, depending on which execution strategy is
used. Ambiguity can be reduced by using a more instantiated strategy class.
The current \chrism{} system uses the refined semantics.
Many programs that are ambiguous in general are unambiguous w.r.t. the refined strategy class,
but not all of them. As a counterexample, consider the program consisting
of the rule
``{\verb|0.5 ?? a, b(X) <=> c(X)|}''
with the query ``{\tt b(1), b(2), a}''.
There are two ways to execute this program in the refined semantics:
one in which the rule instantiation ``{\tt a, b(1)}'' is considered first,
and one in which the rule instantiation ``{\tt a, b(2)}'' is considered first.
According to the first execution strategy, the result is 
``{\tt c(1), b(2)}'' with a probability of 50\%,
``{\tt c(2), b(1)}'' with a probability of 25\%,
and ``{\tt a, b(1), b(2)}'' with a probability of 25\%.
According to the second execution strategy the probabilities of the first two
outcomes are switched.

\paragraph{Ambiguity vs. confluence.}
Ambiguity of \chrism{} programs is related to confluence \cite{abd_fru_meuss_confluence_semantics_csr_constr99}
of CHR programs.
A CHR program is confluent if for every query, all derivations (under the \omegat{} semantics)
lead to equivalent final states.
Confluent CHR programs tend to correspond to unambiguous \chrism{} programs.
For example, programs with only propagation rules are always confluent and unambiguous.
However, confluence and unambiguity do not coincide.
For example, a program consisting of the rule ``{\tt a <=> b:0.5 ; c:0.5}''
is not confluent (because for the query ``{\tt a}'' it has two non-equivalent final states)
but it is unambiguous.
Vice versa, some programs are confluent CHR programs while they are ambiguous \chrism{} programs.
For example, consider the following program:
\begin{verbatim}
   0.5 ?? a <=> b.
   0.5 ?? a <=> c.
   0.5 ?? c <=> b.
\end{verbatim}
If we ignore the probabilities and consider this as a regular CHR program,
then we get a confluent program (all derivations for the query ``{\tt a}'' end in the result ``{\tt b}'').
However, as a \chrism{} program, it is ambiguous.
If the execution strategy is such that the first rule is considered first for the query ``{\tt a}'',
then the probability of ending up with the result ``{\tt b}'' is 67.5\%.
Using an execution strategy that considers the second rule first, the probability of getting ``{\tt b}''
is only 50\%. Therefore, the probability depends on the execution strategy and the program is ambiguous.



\section{Implementation of \Chrism{}}
The implementation of \chrism{} is based on a source-to-source transformation
from \chrism{} rules to CHR(PRISM) rules. 
PRISM is implemented on top of B-Prolog, and several CHR systems
are currently available for B-Prolog.
In \cite{sney_meert_vennekens_chrism_chr09} we
presented a prototype implementation of \chrism{} 
that used a naive CHR(PRISM) system based on {\tt toychr}\footnote{by Gregory J. Duck, 2004.
Download: {\href{http://www.cs.mu.oz.au/~gjd/toychr/}{http://www.cs.mu.oz.au/$\sim$gjd/toychr/}}},
which is a rather naive implementation of (ground) CHR in pure Prolog.
The current implementation of \chrism{}\footnote{
Download the \chrism{} system at
\href{http://people.cs.kuleuven.be/jon.sneyers/chrism/}{http://people.cs.kuleuven.be/jon.sneyers/chrism/}}
is based on the more advanced Leuven CHR system \cite{schr_demoen_kulchr_chr04}.

\subsection{PRISM}
PRISM \cite{prism} is a probabilistic logic programming language.
It is an extension of Prolog with a probabilistic built-in 
\emph{multi-valued random switch (msw)}. 
A multi-valued switch atom {\tt msw(exp, Result)} represents a probabilistic experiment 
named {\tt exp} (a ground Prolog term), which produces an outcome {\tt Result}.  
The set of possible outcomes for such an experiment is defined by means of a 
predicate {\tt values(term,[v1,..., vn])} and {\tt term} unifies with {\tt exp}.  By default, a uniform distribution
is assumed (all values are equally likely).
Different probabilities can be assigned by means of \verb|set_sw(term, [p1, ..., pn])|.

A PRISM program consists out of two parts, 
rules $R$ and facts $F$. The facts $F$ define a {\em base probability distribution} $P_F$ on {\tt msw}-atoms, 
by means of the {\tt values}/2 and \verb|set_sw|/2 predicates.   
The rules $R$ are a set of definite clauses, which are allowed 
to contain the {\tt msw} predicate in the body (but not in the head).  
This set of clauses $R$ serves to extend the base distribution $P$ to a 
distribution $P_{DB}(\cdot)$ over the set of Herbrand interpretations: 
for each interpretation $M$ of the {\tt msw} terms, the probability $P_{F}(M)$ 
is assigned to the interpretation $I$ that is the least Herbrand model 
of $R \cup M$ (\emph{distribution semantics}).

\subsection{Transformation to CHR(PRISM)}
The transformation from \chrism{} to CHR(PRISM) is rather straightforward
and can be done efficiently (linear time).
We illustrate it by example.
Consider again the rule
``\verb|player(P) <=> choice(P) ?? rock(P) ; scissors(P) ; paper(P)|''
from Section~\ref{sec:rsp}. It is translated to the following
CHR(PRISM) code:
\begin{verbatim}
   values(choice(_),[1,2,3]).
   player(P) <=> msw(choice(P),X), 
                 (X=1->rock(P); X=2->scissors(P); X=3->paper(P)).
\end{verbatim}

\noindent
Another example is the random graph rule from Section~\ref{sec:rgraph}:
\begin{verbatim}
   eval(3/(N-1)) ?? nb_nodes(N), node(A), node(B) ==> edge(A,B).
\end{verbatim}
which gets translated to the following CHR(PRISM) code:
\begin{verbatim}
   values(experiment1,[1,2]).
   nb_nodes(N), node(A), node(B) ==> 
        P1 is 3/(N-1), P2 is 1-P1, set_sw(experiment1,[P1,P2]),
        msw(experiment1,X),  (X=1 -> edge(A,B)  ;  X=2 -> true).
\end{verbatim}

Probabilistic simplification rules and simpagation rules are a bit
more tricky since it does not suffice to add a ``nop''-disjunct like above.
The reason is that any removed heads are removed from the constraint store 
as soon as the body is entered, and just reinserting the removed heads 
potentially causes nontermination.
Putting the {\tt msw}-test in the guard of the rule also does not work
as expected. In sampling mode, this works fine, but when doing probability
computations or learning, an unwanted behavior emerges because of
the way PRISM implements explanation search. During explanation search,
PRISM essentially redefines {\tt msw}/2 such that it creates a choice point and
tries all values. This causes the guard to
always succeed and thus explanations that involve \emph{not} firing
a chance rule are erroneously missed.
Hence some care has to be taken to translate such rules to PRISM code
that behaves correctly. 
The solution we have adopted is to add a built-in to CHR to explicitly remove
a constraint from the head of a rule. All \chrism{} rules are translated to
propagation rules. The removed heads are explicitly removed
in the body of the rule, but only in the branch in which the rule instance
is actually applied.

\section{Related Work}


The idea of a probabilistic version of CHR is not new.
In \cite{fru_dipierro_wiklicky_probabilistic_chr_wflp02}, a probabilistic
variant of CHR, called PCHR, was introduced.
In PCHR, every rule gets a weight representing a relative probability.
A rule is chosen randomly from all applicable rules, according to a probability
distribution given by the normalized weights.
For example, the following PCHR program implements a 
coin toss:
\begin{verbatim}
   toss <=>0.5: head.
   toss <=>0.5: tail.
\end{verbatim}

One of the conceptual advantages of PCHR, at least 
from a theoretical point of view, is that its semantics instantiates
the abstract operational semantics $\omegat$ of CHR \cite{newsurvey}: every PCHR derivation
corresponds to some $\omegat$ derivation.

However, the semantics of PCHR may also lead to some confusion, since it is
not so clear what the meaning of the rule weight really is. 
For example, consider again the above coin tossing example. 
For the query {\tt toss} we get the answer {\tt head} with 50\% chance and
otherwise {\tt tail}, so one may be tempted to interpret weights as rule probabilities.
However, if the second rule is removed from the program, we do not get the 
answer {\tt head} with 50\% chance, but with a probability of 100\%. 
The reason is that the weights are normalized w.r.t. the sum of the weights of
all applicable rules.
As a result of this normalization, the actual probability of a rule can only be
computed at runtime and by considering the full program. In other words,
the probabilistic meaning of a single rule heavily depends on the rest of the PCHR program;
there is no localized meaning.
Also, adding weights to \emph{propagation} rules 
is not very useful in practice.
\comment{
Consider for example this program:
\begin{verbatim}
   toss ==>0.5: head.
   toss ==>0.5: tail.
\end{verbatim}
For the above program, the result of the query {\tt toss} will always be
{\tt toss, head, tail} (not necessarily in that order): the first propagation rule
to fire is chosen randomly, and then the other one has to fire because it is the
only applicable rule that is left (so its weight effectively does not matter).
}

The abstract semantics \omegat{} of CHR can be instantiated to allow more execution control and more
efficient implementations.
However, the PCHR semantics, even though it conforms to \omegat{}, 
cannot be instantiated in a similar way.
The reason is that the semantics of PCHR refers to all applicable rules in order to randomly pick one.
This conflicts fundamentally with the purpose of instantiations like the refined semantics,
which consider only a small fragment of the set of applicable rules,
e.g. only rules for the current active constraint occurrence.

The \omegachrism{} semantics of \chrism{} differs from that of PCHR.
In particular, \omegachrism{} derivations do not always correspond
to \omegat{} derivations 
(although they do, in a sense, correspond to {\em unfinished} \omegat{} derivations).
However, the semantics of \chrism{} can be instantiated since 
chance rules have a localized meaning: the application probability does not 
depend on the set of all applicable rules like in PCHR.
As a result, it can be implemented 
efficiently and more execution control can be obtained.

Another advantage of  \chrism{} over PCHR are the features inherited from PRISM,
in particular probability computation and EM-learning.
The existing PCHR implementation only supports probabilistic execution, i.e. sampling.

\paragraph{Probabilistic Logic Programming.}
There are numerous probabilistic extensions of logic programming.
One particular family of such extensions is formed by CP-logic or 
LPADs, ProbLog, ICL, 
and PRISM itself \cite{prism}.  All of these can be encoded in \chrism{}:
in \cite{sney_meert_vennekens_chrism_chr09} we have shown that CP-logic 
(of which ProbLog, ICL, etc. are sublogics)
can be encoded 
in \chrism\ in a compact and modular way.


Next to these ``logic programming flavored'' languages, there are also a number 
of formalisms that are inspired by Bayesian networks, such as BLP, 
RBN, CLP(BN), and Blog.  
Based on the encoding of Bayesian networks that we gave in Section \ref{sec:bn}, 
we can also translate BLPs to \chrism.  
RBNs, CLP(BN) and Blog would be more difficult, 
because they allow more complex probability distributions, for which \chrism\ currently 
does not offer support. (A more detailed description of these formalisms can be found in \cite{Getoor07}.)

\comment{
As the above paragraphs show, we can encode CP-logic or BLPs using only the \verb|D1:P1; ...; Dn:Pn| construction of \chrism.  In particular, the ability to assign a probability to an entire rule is not used.  This construct offers some expressivity that these other language do not have; namely, it allows to explicitly make things false.   For instance, suppose we want to model the evolution of a population.  The fact that an individual might die could be represented in \chrism\ as:
\verb|0.1 ?? alive <=> true|.
In a language such as CP-logic, this would have to be encoding in a roundabout way, using explicit time points and a probabilistic law of persistence:
\[ (alive(T+1) :0.9) \leftarrow alive(T).\]
The \chrism\ representation is more compact and elegant.
}


\section{Potential Applications}

Both PRISM and CHR have been successfully applied in a wide range of research fields. 
Since the features of PRISM and CHR are largely orthogonal, we can expect \chrism{} 
to be extremely suitable for applications at the intersection of the application 
areas of PRISM and CHR. One example of an application area at the intersection is 
abduction, which has been studied in the context of PRISM \cite{prism_abduction} and also in the 
context of CHR (\citeN{newsurvey}, Section 7.3.2). Computational linguistics  and 
bio-informatics  are two other domains in which both PRISM and CHR 
have proven to be very valuable tools
\cite{prism,christ_chr_grammars_tplp05,prism_bio}.

Furthermore, in many application domains of CHR, 
there is clearly a potential for probabilistic extensions of the existing approaches, 
for instance to deal with uncertain information.
Examples are (section numbers refer to \citeN{newsurvey}):
scheduling (Section 7.1.1), spatio-temporal reasoning and robotics 
(Section 7.1.2), multi-agent systems (Section 7.1.3), 
the semantic web (Section 7.1.4), type systems (Section 7.3.1), 
testing and verification (Section 7.3.5).

Another interesting application area is the automatic analysis and generation of music. 
In the past, we have used PRISM to analyse and generate music in a probabilistic setting \cite{prism_music}. 
There are also several deterministic approaches based on constraints and strict rules (e.g. \citeN{asp_music}). 
Preliminary results indicate that a combined approach, using CHRiSM, is very promising. 
In this application, sampling of a probabilistic model corresponds to music generation, 
while parameter learning from a training set corresponds to tuning the model to a specific 
genre or composer, and probability computation (or Viterbi computation) can be used for music classification.

\section{Conclusion}

In this exploratory paper, we have introduced a novel rule-based
probabilistic-logic formalism called \chrism{}, which is based on
a combination of CHR and PRISM. 
We have defined an operational semantics for arbitrary \chrism{} programs
and a distribution semantics for unambiguous \chrism{} programs.
We have illustrated the \chrism{} system by example and we have
outlined some potential application areas in which \chrism{} can be used.
Finally, we have sketched the implementation of the \chrism{} system
and discussed related languages, in particular PCHR.

In our opinion, CHR has important advantages over Prolog,
including complexity-wise completeness and the expressivity of multi-headed rules.
We expect \chrism{} to have the same advantages over plain PRISM.


There are several directions for future work.
The notion of ambiguity and its relation to confluence has to be explored;
in particular, the existence of a decidable ambiguity test (for terminating \chrism{} programs).
Although the current implementation is sufficiently efficient for sampling,
it is too naive for probability computation and learning, since those tasks
require an efficient mechanism to find explanations (sequences of probabilistic choices)
for observations. 
Improving the efficiency of explanation search is the topic of ongoing work
\cite{result-directed}.
Another limitation of the current implementation is that it only supports ground
queries and observations.
Finally, it would be interesting to transfer automatic CHR program generation techniques
(e.g. \citeN{abd_et_al_arm_lopstr06}) to \chrism{} in order to obtain
a system that supports not only parameter learning but also structure learning (rule learning).

\bibliographystyle{acmtrans} 
\bibliography{biblio}

\end{document}